\documentclass[12pt]{article}

\usepackage{amssymb}
\usepackage{amsmath}
\usepackage{epsfig}

\begin{document}

\title{\bf Effects of Schwarzschild Geometry on Isothermal Plasma Wave Dispersion}

\author{M. Sharif \thanks{msharif@math.pu.edu.pk} and Umber Sheikh\\
{\small Department of Mathematics,}\\
{\small University of the Punjab, Lahore 54590, Pakistan}}

\date{\small{Dated:04-12-07}}

\maketitle

\begin*{}
The behavior of isothermal plasma waves has been analyzed near the
Schwarzschild horizon. We consider a non-rotating background with
non-magnetized and magnetized plasmas. The general relativistic
magnetohydrodynamical equations for the Schwarzschild planar
analogue spacetime with an isothermal state of the plasma are
formulated. The perturbed form of these equations is linearized and
Fourier analyzed by introducing simple harmonic waves. The
determinant of these equations in each case leads to a complex
dispersion relation, which gives complex values of the wave number.
This has been used to discuss the nature of the waves and their
characteristics near the horizon.
\end*{}
\pagebreak

\section{INTRODUCTION}

General Relativity (GR) is a beautiful scheme for describing the
gravitational field. This theory is believed to apply to all forms
of interactions, especially between large scale gravitational
structures. It has been proven that black holes exist on the basis
of study of the effects they exert on their surroundings. They
greatly affect the surrounding plasma medium with their enormous
gravitational fields. Since all compact objects have strong
gravitational fields near their surfaces \cite{P}, it is important
to study the general relativistic effects on physical processes,
like electromagnetic processes taking place in their vicinity.
Magnetohydrodynamics (MHD) with the effects of gravity is called
general relativistic magnetohydrodynamics (GRMHD). The GRMHD
equations help us to study stationary configurations and dynamic
evolution of a conducting fluid in a magnetosphere. They include
Maxwell's equations, Ohm's law and mass, momentum, and energy
conservation equations. These equations are required to
investigate various aspects of the interaction of relativistic
gravity with a plasma's magnetic field.

The 3+1 formalism is well-suited to carry non-relativistic intuition
of physicists about plasmas, hydrodynamics, and stellar dynamics
into the arena of black holes and general relativity. This formalism
(also called the ADM formalism) was originally developed by Arnowitt
et al. \cite{ADM}. It was motivated by several startling results
proved in the 1970s by using a black-hole viewpoint [3-9]. Thorne
and Macdonald \cite{TM} and Thorne et al. \cite{TPM} extended this
formulation to electromagnetic fields in black-hole theory. The wave
propagation theory in Friedmann universe was investigated in a 3+1
formalism by Holcomb and Tajima \cite{HT}, Holcomb \cite{Ho}, and
Dettmann et al. \cite{De}. Khanna \cite{Kh} derived the MHD
equations describing the two-component plasma theory of a Kerr black
hole in this split. Komissarov \cite{Ko1} discussed the
Blandford-Znajek monopole solution by using the 3+1 formalism in
black-hole electrodynamics. Zhang [17,18] formulated the black-hole
theory for stationary symmetric GRMHD by using the 3+1 formalism
with its applications on a Kerr geometry.

The formalism for gravitational perturbations away from the
Schwarzschild background has been developed by Regge and Wheeler
\cite{Re}. It was extended by Zerilli \cite{Ze}, who showed that the
perturbations corresponding to changes in the mass, the angular
momentum, and the charge of the Schwarzschild black hole are
well-behaved. The decay of non-well-behaved perturbations has been
investigated by Price \cite{Pr}. The quasi-static electric problem
was solved by Hanni and Ruffini \cite{HR}, who proved that the lines
of force diverge at the horizon for an observer at infinity. Wald
\cite{Wa} derived the solution for the electromagnetic field
occurring when a stationary axisymmetric black hole is placed in a
uniform magnetic field aligned along the symmetry axis of the black
hole. A linearized treatment of plasma waves for a special
relativistic formulation of the Schwarzschild black hole was
developed by Sakai and Kawata \cite{SK}. Buzzi et al. \cite{BH}
extended this treatment to waves propagating in the radial direction
in a general-relativistic two-component plasma by using the 3+1 ADM
formalism. They investigated wave modes for one-dimensional radial
propagation of transverse and longitudinal waves close to the
Schwarzschild horizon.

The aim of this paper is to investigate the properties of isothermal
plasma waves by taking the Schwarzschild spacetime in a planar
analogue. The scheme of the paper is as follows: The next section
provides comprehensive details about the results in the 3+1
formalism. In Section III, we obtain the GRMHD equations for a
perfect fluid with perfect MHD flow conditions. Section IV describes
the perturbations and relative assumptions that will be used in the
ensuing sections. In Section V, we apply the perturbations and
Fourier-analysis methods to the GRMHD equations. Section VI
investigates the dispersion relations obtained in graphical forms.
The last section contains a summary and discussion of the results.

\section{3+1 SPACETIME MODELING}

In the $3+1$ formalism, the line element of the spacetime can be
written as \cite{Z2}
\begin{equation}\label{a}
ds^2=-\alpha^2dt^2+\gamma_{ij}(dx^i+\beta^idt)(dx^j+\beta^jdt),
\end{equation}
where $\alpha$ is the lapse function, $\beta^i$ are components of
the shift vector, and $\gamma_{ij}$ are components of the spatial
metric. All these quantities are functions of coordinates $t$ and
$x^i$. A natural observer, associated with this spacetime and
called the fiducial observer (FIDO), has a four-velocity
\textbf{n} perpendicular to the hypersurfaces of constant time $t$
and is given by
\begin{equation*}
\bf{n}=\frac{1}{\alpha}(\frac{\partial}{\partial
t}-\beta^i\frac{\partial}{\partial x^i}).
\end{equation*}
Notice that we are using geometrized units so that $G=1=c$. In
four-dimensional spacetime, vectors and tensors will be denoted by
boldface italic letters. A letter having a dyad over it will show
a three-dimensional tensor. All vector analysis notations, such as
gradient, curl and vector product, will be those of the
three-dimensional absolute space with the three metric
$\overleftrightarrow{\gamma}$. The Latin letters $i, j, k,...$
represent indices in absolute space and run from 1 to 3.

The perfect MHD flow assumption \cite{Z1} is
\begin{equation}\label{a5}
\textbf{E}+\textbf{V}\times\textbf{B}=0,
\end{equation}
where \textbf{V}, \textbf{E}, and \textbf{B} are the velocity and
the electric and magnetic fields of the fluid as measured by the
FIDO. This shows that there can be no electric field in the fluid's
rest-frame.

Applying this condition, Faraday's law in the 3+1 formalism
becomes
\begin{equation}\label{a6}
\frac{d\textbf{B}}{d\tau}+\frac{1}{\alpha}(\textbf{B}.\nabla)\beta
+\theta\textbf{B}=\frac{1}{\alpha}\nabla\times(\alpha\textbf{V}\times\textbf{B}),
\end{equation}
where the FIDO's four-velocity expansion rate and FIDO's measured
rate of change of any three-dimensional vector in absolute space
(i.e., orthogonal to \textbf{n}) can be expressed, respectively,
as
\begin{equation*}
\theta=\frac{1}{\alpha}(\frac{g_{,t}}{2g}-\nabla.\beta),\quad
\frac{d}{d \tau}\equiv \frac{1}{\alpha}(\frac{\partial}{\partial
t}-\beta.\nabla).
\end{equation*}
For perfect MHD flow, the equations of evolution of the magnetic
field give
\begin{equation}\label{a7}
\frac{D\textbf{B}}{D\tau}+\frac{1}{\alpha}(\textbf{B}.\nabla)(\beta-\alpha\textbf{V})
+\left\{\theta+\frac{1}{\alpha}\nabla.(\alpha\textbf{V})\right\}\textbf{B}=0,
\end{equation}
where
\begin{equation*}
\frac{D}{D\tau}\equiv\frac{d}{d\tau}+\textbf{V}.\nabla
=\frac{1}{\alpha}\left\{\frac{\partial}{\partial
t}+(\alpha\textbf{V}-\beta).\nabla\right\}
\end{equation*}
is the time derivative moving with the fluid. The local
conservation law of the rest mass according to the FIDO is
\begin{equation}\label{a8}
\frac{D\rho_0}{D\tau}+\rho_0\gamma^2\textbf{V}.\frac{D\textbf{V}}{D\tau}+
\frac{\rho_0}{\alpha}\left\{\frac{g_{,t}}{2g}+\nabla.(\alpha\textbf{V}-\beta)\right\}=0,
\end{equation}
with $\rho_0$ being the rest-mass density. The law of force
balance measured by the FIDO is given by
\begin{eqnarray}\label{a9}
&&\left\{\left(\rho_0\gamma^2\mu+\frac{\textbf{B}^2}{4\pi}\right)\gamma_{ij}+\rho_0\gamma^4\mu
V_iV_j-\frac{1}{4\pi}B_iB_j\right\}\frac{DV^j}{D\tau}+\rho_0\gamma^2V_i\frac{D\mu}{D\tau}\nonumber\\
&&-\left(\frac{\textbf{B}^2}{4\pi}\gamma_{ij}-\frac{1}{4\pi}B_iB_j\right){V^j}_{|k}V^k
=-\rho_0\gamma^2\mu\left\{a_i
-\frac{1}{\alpha}\beta_{j|i}V^j-(\pounds_t\gamma_{ij})V^j\right\}
\nonumber\\
&&-p_{|i}
+\frac{1}{4\pi}(\textbf{V}\times\textbf{B})_i\nabla.(\textbf{V}\times
\textbf{B})-\frac{1}{8\pi\alpha^2}(\alpha\textbf{B})^2_{|i}+\frac{1}{4\pi\alpha}(\alpha
\textbf{B}_i)_{|j}B^j\nonumber\\
&&-\frac{1}{4\pi\alpha}[\textbf{B}\times\{\textbf{V}
\times(\nabla\times(\alpha\textbf{V}\times\textbf{B})
-(\textbf{B}.\nabla)\beta)+(\textbf{V}\times\textbf{B}).\nabla\beta\}]_i,
\end{eqnarray}
where $p$ is the pressure. The FIDO's measured local energy
conservation law \cite{Z1} is given by
\begin{equation}\label{a10}
\frac{d\epsilon}{d\tau}+\theta
\epsilon+\frac{1}{2\alpha}W^{ij}(\pounds_t \gamma^{ij})
=-\frac{1}{\alpha^2}\nabla.(\alpha^2
\bf{S})+\frac{1}{\alpha}\nabla\beta:
\overleftrightarrow{\bf{W}}+\bf{E}.\bf{j}.
\end{equation}
When we substitute the following values for a perfect fluid,
\begin{eqnarray*}
\epsilon&=&(\mu \rho_0-p(1-\textbf{V}^2))\gamma^2,\\
\textbf{S}&=&\mu\rho_0\gamma^2\textbf{V},\\
\overleftrightarrow{\textbf{W}}&=&\mu\rho_0\gamma^2\textbf{V}
\otimes\textbf{V}+p\overleftrightarrow{\gamma},
\end{eqnarray*}
with $\mu$ being the specific enthalpy, and perfect MHD flow
assumption, Eq. (\ref{a10}) becomes
\begin{eqnarray}\label{a11}
&&\rho_0\gamma^2\frac{D\mu}{D\tau}+\mu\gamma^2\frac{D\rho_0}{D\tau}
+2\rho_0\mu\gamma^4\textbf{V}.\frac{D\textbf{V}}{D\tau}-\frac{d
p}{d\tau}+(\mu\rho_0\gamma^2-p)\theta+2\rho_0\mu\gamma^2\textbf{V}.\textbf{a}\nonumber\\
&& +\rho_0\mu\gamma^2(\nabla.\textbf{V})
-\frac{1}{\alpha}\{\rho_0\mu\gamma^2\textbf{V}.(\textbf{V}.\nabla)\beta
+p(\nabla.\beta)\}+\frac{1}{2\alpha}\{\rho_0
\mu\gamma^2V^iV^j\nonumber\\
&&+p\gamma^{ij}\}\pounds_t\gamma_{ij}+\frac{1}{4\pi\alpha}[(\textbf{V}\times\textbf{B}).(\nabla
\times (\alpha\textbf{B}))+\theta\alpha(\textbf{V}\times\textbf{B})\nonumber\\
&&+\alpha(\textbf{V}\times\textbf{B}).\frac{d}{d\tau}(\textbf{V}\times\textbf{B})
+(\textbf{V}\times\textbf{B}).(\textbf{V} \times
\textbf{B}.\nabla)\beta. (\textbf{V}\times\textbf{B})]=0
\end{eqnarray}
with acceleration $a=\frac{\nabla \alpha}{\alpha}$. Equations
(\ref{a6})-(\ref{a9}) and (\ref{a11}) are the perfect GRMHD
equations for a plasma in the vicinity of a black hole.

\section{GRMHD EQUATIONS FOR ISOTHERMAL PLASMA IN SCHWARZSCHILD
PLANAR ANALOGUE}

This section will give a detailed structure of the chosen
spacetime with the respective form of the GRMHD equations.

\subsection{Description of the Planar Schwarzschild Spacetime and the Equation of State}

The planar approximation of the Schwarzschild line element
\cite{Z2} is
\begin{eqnarray}{\setcounter{equation}{1}}\label{p}
ds^2=-\alpha^2(z)dt^2+dx^2+dy^2+dz^2,
\end{eqnarray}
which is an analogue of the Schwarzschild spacetime with $z$, $x$
and $y$ as radial $r$, axial $\phi$, and poloidal $\theta$
directions, respectively. The lapse function vanishes at the
horizon, which can be placed at $z=0$, and increases monotonically
to unity as $z$ increases from $0$ to $\infty$. The relationship
between the universal time $t$ and the FIDO's time $\tau$ can be
expressed by the time lapse $\alpha\equiv\frac{d\tau}{dt}$. The
Schwarzschild black hole is non-rotating; hence, there is no shift
of coordinates. In this spacetime, the  FIDO has a four-velocity
equivalent to $\textbf{n}=\frac{1}{\alpha}\frac{\partial}{\partial
t}$.

In the hydrodynamic treatment, the plasma is represented as a
perfect fluid. In a planar analogue of the Schwarzschild
spacetime, we assume that the system of perfect GRMHD equations is
enclosed by the isothermal state of a plasma. This state can be
expressed by the following equation \cite{Z1}:
\begin{equation}
\mu=\frac{\rho+p}{\rho_0}=constant,
\end{equation}
which shows that there is no exchange of the heat between the
plasma and the magnetic field of fluid.

\subsection{Respective GRMHD Equations}

For the Schwarzschild planar analogue given in Eq. (\ref{p}), the
GRMHD equations Eqs. (\ref{a6})-(\ref{a9}) and (\ref{a11}), in the
3+1 formalism for the isothermal state of a plasma take the form
\begin{eqnarray}\label{b2}
&&\frac{\partial \textbf{B}}{\partial t}=\nabla \times(\alpha
\textbf{V} \times \textbf{B}),\\ \label{b3} &&\frac{\partial
\textbf{B}}{\partial t}+(\alpha
\textbf{V}.\nabla)\textbf{B}-(\textbf{B}.\nabla)(\alpha
\textbf{V})+\textbf{B}\nabla.(\alpha \textbf{V})=0,\\\label{b4}
&&\frac{\partial (\rho+p)}{\partial t}+(\alpha
\textbf{V}.\nabla)(\rho+p)
+(\rho+p)\gamma^2\textbf{V}.\frac{\partial \textbf{V}}{\partial
t}\nonumber\\
&&+(\rho+p)\gamma^2\textbf{V}.(\alpha
\textbf{V}.\nabla)\textbf{V}+(\rho+p)
\nabla.(\alpha\textbf{V})=0,\\
\label{b5}
&&\left\{\left((\rho+p)\gamma^2+\frac{\textbf{B}^2}{4\pi}\right)\delta_{ij}
+(\rho+p)\gamma^4V_iV_j-\frac{1}{4\pi}B_iB_j\right\}
\left(\frac{1}{\alpha}\frac{\partial}{\partial
t}+\textbf{V}.\nabla\right)V^j\nonumber\\
&&-\left(\frac{\textbf{B}^2}{4\pi}\delta_{ij}-\frac{1}{4 \pi}B_i
B_j\right)V^j_{,k}V^k=-(\rho+p)\gamma^2 a_i+\frac{1}{4\pi}
\nabla.(\textbf{V}\times\textbf{B})(\textbf{V}\times \textbf{B})_i\nonumber\\
&&-\frac{(\alpha \textbf{B})^2_{,i}}{8\pi\alpha^2}+\frac{(\alpha
B_i)_{,j}B^j}{4 \pi \alpha}-\frac{1}{4 \pi
\alpha}[\textbf{B}\times\{\textbf{V} \times (\nabla \times (\alpha
\textbf{V} \times \textbf{B}))\}]_i-p_{,i},\\
\label{b6} &&\gamma^2(\frac{1}{\alpha}\frac{\partial}{\partial
t}+\textbf{V}.\nabla)(\rho+p)
+2(\rho+p)\gamma^4\textbf{V}.(\frac{1}{\alpha}\frac{\partial}{\partial
t}+\textbf{V}.\nabla)\textbf{V}\nonumber
\end{eqnarray}
\begin{eqnarray}
&&+2(\rho+p)\gamma^2\textbf{V}.\textbf{a}
-\frac{1}{\alpha}\frac{\partial p}{\partial
t}+(\rho+p)\gamma^2\nabla.\textbf{V}\nonumber\\
&&+\frac{1}{4\pi\alpha}[(\textbf{V} \times \textbf{B}).(\nabla
\times (\alpha \textbf{B}))+ \alpha
(\textbf{V}\times\textbf{B}).\frac{1}{\alpha}
\frac{\partial}{\partial t}(\textbf{V} \times \textbf{B})]=0.
\end{eqnarray}
Equations (\ref{b2})-(\ref{b6}) are the perfect GRMHD equations for
an isothermal plasma in the vicinity of a Schwarzschild black hole.
In the rest of the paper, we shall analyze these equations by using
perturbations and Fourier analysis procedures.

\section{RELATIVE ASSUMPTIONS}

In this background, the perturbed flow of the fluid is considered
to be along the $z$-axis. Thus the FIDO's measured four-velocity
and magnetic field are along the $z$-axis and can be expressed as
\begin{equation*}
\textbf{V}=u(z)\textbf{e}_\textbf{z},\quad\textbf{B}=B\textbf{e}_\textbf{z},
\end{equation*}
where $B$ is a constant quantity. For this velocity, the Lorentz
factor $\gamma$ takes the form $\gamma=\frac{1}{\sqrt{1-u^2}}.$

The perturbed flow in the magnetosphere shall be characterized by
its fluid density $\rho$, pressure $p$, velocity \textbf{V}, and
magnetic field \textbf{B} (as measured by the FIDO). The
first-order perturbations in the above-mentioned quantities are
denoted by $\delta \rho,~\delta p,~\delta\textbf{V}$, and $\delta
\textbf{B}$. Consequently, the perturbed variables will take the
following form:
\begin{eqnarray}{\setcounter{equation}{1}}\label{b7}
\rho=\rho^0+\delta{\rho}=\rho^0+\rho \tilde{\rho},&~&
p=p^0+\delta{p}=p^0+p \tilde{p},\nonumber\\
\textbf{B}=\textbf{B}^0+\delta{\textbf{B}}=\textbf{B}^0+B\textbf{b},&~&
\textbf{V}=\textbf{V}^0+\delta{\textbf{V}}=\textbf{V}^0+\textbf{v},
\end{eqnarray}
where $\rho^0,~p^0,~\textbf{B}^0$ and $\textbf{V}^0$ are unperturbed
quantities. Due to gravitation, the waves can propagate in the
$z$-direction with respect to time $t$. Hence, the perturbed
quantities depend on $z$ and $t$.

The following notations for the perturbed quantities will be used:
\begin{eqnarray*}
\tilde{\rho}\equiv\frac{\delta \rho}{\rho}=\tilde{\rho}(t,z),&~&
\tilde{p}\equiv\frac{\delta p}{p}=\tilde{p}(t,z),\\
\textbf{v}\equiv \delta \textbf{V}=v_z(t,z)\textbf{e}_\textbf{z},&~&
\textbf{b}\equiv \frac{\delta \textbf{B}}{B}
=b_z(t,z)\textbf{e}_\textbf{z}.
\end{eqnarray*}
We also assume that the perturbations have harmonic space and time
dependences, i.e.,
\begin{equation*}
\tilde{\rho},~\tilde{p},~\textbf{v}~\rm{and}~\textbf{b}\sim
\it{e^{-i (\omega t-k z)}}.
\end{equation*}
Thus, the perturbed variables can be expressed as
\begin{eqnarray}\label{c1}
\tilde{\rho}(t,z)=c_1e^{-\iota(\omega t-kz)},\nonumber\\
\tilde{p}(t,z)=c_2e^{-\iota(\omega t-kz)},\nonumber\\
v_z(t,z)=c_3e^{-\iota (\omega t-kz)},\nonumber\\
b_z(t,z)=c_4e^{-\iota (\omega t-kz)}.
\end{eqnarray}

\section{PERTURBATION EQUATIONS}

If the perturbations given by Eq. (\ref{b7}) are introduced, the
GRMHD equations, Eqs. (\ref{b2})-(\ref{b6}), become
\begin{eqnarray}{\setcounter{equation}{1}}\label{c2}
&&\frac{\partial(\delta \textbf{B})}{\partial t}=\nabla
\times(\alpha\textbf{v}\times\textbf{B})
+\nabla\times(\alpha\textbf{V}\times\delta\textbf{B}),\\
&&\nabla.(\delta\textbf{B})=0,\\
&&\frac{\partial(\delta\rho+\delta p)}{\partial
t}+(\alpha\textbf{V}.\nabla)(\delta\rho+\delta
p)+(\rho+p)\gamma^2\textbf{V}.\frac{\partial\textbf{v}}{\partial
t}-\alpha(\rho+p)\textbf{v}.\nabla \ln u\nonumber\\&&
+\alpha(\rho+p)(\nabla.\textbf{v})+(\delta\rho+\delta
p)\nabla.(\alpha\textbf{V})+(\delta\rho+\delta p)\gamma^2
\textbf{V}.(\alpha\textbf{V}.\nabla)\textbf{V}\nonumber\\
&&+2(\rho+p)\gamma^2(\textbf{V}.\textbf{v})(\alpha\textbf{V}.\nabla)
\ln\gamma+(\rho+p)\gamma^2(\alpha\textbf{V}.\nabla\textbf{V}).\textbf{v}\nonumber\\
&&+(\rho+p)\gamma^2\textbf{V}.(\alpha\textbf{V}.\nabla)\textbf{v}=0,\\
&&\{((\rho+p)\gamma^2+\frac{\textbf{B}^2}{4\pi})\delta_{ij}+(\rho+p)\gamma^4
V_iV_j-\frac{1}{4\pi}B_iB_j\}\frac{1}{\alpha}\frac{\partial
v^j}{\partial
t}\nonumber\\
&&+\frac{1}{4\pi}[\textbf{B}\times\{\textbf{V}\times\frac{1}{\alpha}
\frac{\partial(\delta\textbf{B})}{\partial
t}\}]_i+(\rho+p)\gamma^2v_{i,j}V^j+(\rho+p)\gamma^4V_iv_{j,k}V^jV^k\nonumber\\
&&-\frac{1}{4\pi\alpha} \{(\alpha\delta B_i)_{,j}-(\alpha\delta
B_j)_{,i}\}B^j=-\gamma^2\{(\delta\rho+\delta
p)+2(\rho+p)\gamma^2(\textbf{V}.\textbf{v})\}a_i\nonumber\\
&&-(\delta p)_{,i}+\frac{1}{4\pi\alpha}\{(\alpha B_i)_{,j}
-(\alpha B_j)_{,i}\}\delta B^j-(\rho+p)\gamma^4 (v_iV^j+v^jV_i)V_{k,j}V^k\nonumber\\
&&-\gamma^2\{(\delta\rho+\delta p)V^j
+2(\rho+p)\gamma^2(\textbf{V}.\textbf{v})V^j+(\rho+p)v^j\}V_{i,j}\nonumber\\
&&-\gamma^4V_i\{(\delta\rho+\delta
p)V^j+4(\rho+p)\gamma^2(\textbf{V}.\textbf{v})V^j+(\rho+p)v^j\}V_{j,k}V^k,\\
\label{c3} &&\gamma^2\frac{1}{\alpha}\frac{\partial
(\delta\rho+\delta p)}{\partial
t}+\frac{2}{\alpha}(\rho+p)\gamma^4\textbf{V}.\frac{\partial\textbf{v}}{\partial
t}-2(\rho+p)\gamma^4(\textbf{V}.\textbf{v})(\textbf{V}.\nabla)\ln
u\nonumber\\
&&+6(\rho+p)\gamma^6(\textbf{V}.\textbf{v})\{\textbf{V}.(\textbf{V}.\nabla)\textbf{V}\}
+(\rho+p)\gamma^4\textbf{V}.(\textbf{v}.\nabla)\textbf{V}
\nonumber\\
&&+2(\rho+p)\gamma^4
\textbf{V}.(\textbf{V}.\nabla)\textbf{v}+2(\delta\rho+\delta
p)\gamma^2\textbf{V}.\textbf{a}+2(\rho+p)\gamma^4(\textbf{V}.\textbf{v})
(\textbf{V}.\textbf{a}) \nonumber\\
&&+(\rho+p)\gamma^2(\nabla.\textbf{v})+(\rho+p)\gamma^2\textbf{v}.\textbf{a}
+\gamma^2(\textbf{V}.\nabla)(\delta\rho+\delta p)
-\frac{1}{\alpha}\frac{\partial(\delta p)}{\partial t}\nonumber\\
&&-(\rho+p)\gamma^2 (\textbf{v}.\nabla)\ln u+2(\delta \rho+\delta
p)\gamma^4
\textbf{V}.(\textbf{V}.\nabla)\textbf{V}+(\delta\rho+\delta
p)\gamma^2(\nabla.\textbf{V})\nonumber\\
&&+2(\rho+p)\gamma^4(\textbf{V}.\textbf{v})(\nabla.\textbf{V})
+2(\rho+p)\gamma^4\textbf{v}.(\textbf{V}.\nabla)\textbf{V}
+\frac{1}{4\pi\alpha}[(\textbf{V}\times\textbf{B}).
\nonumber\\
&&(\nabla\times(\alpha\delta\textbf{B}))+
(\textbf{v}\times\textbf{B}).(\nabla\times(\alpha\textbf{B}))+(\textbf{V}
\times\delta\textbf{B}).(\nabla \times (\alpha\textbf{B}))\nonumber\\
&&+(\textbf{V}\times \textbf{B}).(\textbf{V}\times \frac{\partial
\delta\textbf{B}}{\partial t})
+(\textbf{V}\times\textbf{B}).(\frac{\partial\textbf{v}}{\partial
t}\times\textbf{B})]=0.
\end{eqnarray}
The component forms of Eqs. (\ref{c2})-(\ref{c3}) are given as
\begin{eqnarray}\label{c4}
&&\frac{1}{\alpha}\frac{\partial b_z}{\partial t}=0,\\
&&b_{z,z}=0,\\ \label{c5} &&\rho\frac{\partial
\tilde{\rho}}{\partial
t}+u\alpha(\rho\tilde{\rho}_{,z}+p\tilde{p}_{,z})+p\frac{\partial
\tilde{p}}{\partial t}+\gamma^2u(\rho+p)\frac{\partial v_z}{\partial
t}\nonumber\\
&&+\alpha(\rho+p)(1+\gamma^2
u^2)v_{z,z}-(\tilde{\rho}-\tilde{p})\{(\alpha u p)'+\alpha \gamma^2
u^2 p
u'\}\nonumber\\
&& =\alpha(\rho+p)(1-2\gamma^2u^2)(1+\gamma^2
u^2)\frac{u'}{u}v_z,\\
&&(\rho+p)\gamma^2(1+\gamma^2 u^2)\left\{\frac{1}{\alpha}
\frac{\partial v_z}{\partial t}+u
v_{z,z}\right\}+p\tilde{p}_{,z}+\tilde{p}p_{,z}\nonumber\\
&&=-\gamma^2(\rho+p)\{u'(1+\gamma^2 u^2)(1+4\gamma^2 u^2)
+2u\gamma^2a_z)\}v_z\nonumber\\
&&-(\rho\tilde{\rho}+p\tilde{p})\gamma^2\{a_z+u(1+\gamma^2u^2)u'\},\\\label{c6}
&&\rho\gamma^2\frac{1}{\alpha}\frac{\partial}{\partial t}
\tilde{\rho}+p\gamma^2\frac{1}{\alpha}\frac{\partial}{\partial
t}\tilde{p}+\frac{2}{\alpha}(\rho+p)\gamma^4u\frac{\partial
v_z}{\partial t}+(\rho+p)\gamma^2(1+2\gamma^2u^2)v_{z,z}
\nonumber\\
&&+(\rho\tilde{\rho}+p\tilde{p})\gamma^2\{2ua_z+(1+2\gamma^2u^2)u'\}+\gamma^2u(\rho_{,z}\tilde{\rho}
+\tilde{\rho}_{,z}\rho+p_{,z}\tilde{p}+\tilde{p}_{,z}p)
\nonumber\\
&&-p\frac{1}{\alpha}\frac{\partial\tilde{p}}{\partial
t}+(\rho+p)\gamma^2\left\{(3\gamma^2uu'+a_z)(1+2\gamma^2 u^2)
-\frac{u'}{u}\right\}v_z=0.
\end{eqnarray}
The conservation law of rest-mass \cite{Z1} in a three-dimensional
hypersurface for the isothermal state of a plasma,
$$\alpha (\rho+p)\gamma u=constant,$$ is used to obtain Eqs. (\ref{c5})
and (\ref{c6}). The same law will be applied for further
simplifications.

Using Eq. (\ref{c1}) in Eqs. (\ref{c4})-(\ref{c6}), we obtain the
Fourier-analyzed form of the above equations as
\begin{eqnarray}\label{c7}
&&-\frac{\iota \omega}{\alpha} c_4=0,\\\label{c8} &&\iota k
c_4=0,\\\label{c9} &&c_1\{-\rho\iota \omega+\iota k\rho\alpha
u-(u\alpha p)'-\alpha u^2\gamma^2pu'\}+c_2\{-p\iota \omega+\iota kp\alpha u\nonumber\\
&&+(u\alpha p)'+\alpha
u^2\gamma^2pu'\}+c_3(\rho+p)\left\{\alpha(1+\gamma^2u^2)\iota
k\right.\nonumber\\
&&\left.-\alpha(1-2\gamma^2u^2)(1+\gamma^2u^2)\frac{u'}{u}-i\omega\gamma^2 u\right\}=0,\\
&&c_1\rho\gamma^2 \{a_z+uu'(1+\gamma^2 u^2)\}+c_2\{p\gamma^2
\{a_z+uu'(1+\gamma^2 u^2)\}+\iota k p+p'\}\nonumber\\
&&+c_3(\rho+p)\gamma^2\left[(1+\gamma^2u^2)\left(\frac{-\iota
\omega}{\alpha}+\iota u k\right)+\{u'(1+\gamma^2 u^2)(1+4\gamma^2
u^2)\right.\nonumber\\
&&\left.+2ua_z\}\right]=0,\\\label{c10}
&&\left\{\rho\gamma^2\left(\frac{-\iota\omega}{\alpha}+\iota
ku\right)
+\gamma^2u\rho'+\rho2\gamma^2ua_z+\rho(1+2\gamma^2u^2)\gamma^2u'\right\}c_1\nonumber\\
&&+\left\{\frac{-\iota\omega}{\alpha}p(\gamma^2-1)+\iota
k\gamma^2up+\gamma^2up'+2\gamma^2upa_z+p(1+2\gamma^2u^2)\gamma^2u'\right\}c_2\nonumber\\
&&+(\rho+p)\gamma^2\left\{\frac{-2\iota\omega}{\alpha}\gamma^2u+\iota
k(1+2\gamma^2u^2)+3\gamma^2uu'(1+2\gamma^2u^2)\right.\nonumber\\
&&\left.+a_z(1+2\gamma^2u^2)-\frac{u'}{u}\right\}c_3=0.
\end{eqnarray}
Equations (\ref{c7}) and (\ref{c8}) yield that $c_4=0$; i.e., the
magnetic field is not affected by the black-hole gravity and time.
It is mentioned here that Eqs. (\ref{c9})-(\ref{c10}) also give a
non-magnetized plasma.

\section{NUMERICAL SOLUTIONS}

We consider that the magnetosphere is filled with a stiff fluid
for which $\rho=constant=p$ and let $\alpha=\tanh(z)$. Using these
values, the mass conservation law in 3-dimensions gives
$u=\frac{1}{\sqrt{1+\tanh^2(z)}}$.

When we use these values, the determinant of the coefficients of
constants $c_1,~c_2$, and $c_3$ in Eqs. (\ref{c9})-(\ref{c10})
give a complex dispersion equation of the type
\begin{eqnarray}\setcounter{equation}{1}
&&A_1(z)k^2+A_2(z,\omega)k+A_3(z,\omega)+\iota\{A_4(z)k^3\nonumber\\
&&+A_5(z,\omega)k^2+A_6(z,\omega)+A_7(z,\omega)k\}=0,
\end{eqnarray}
which, on solving, gives three complex values of $k=k_1+\iota k_2$.
The sinusoidal expressions then take the form $e^{-\iota(\omega
t-k_1 z-\iota k_2z)}=e^{-\iota(\omega t-k_1 z)-k_2z}$. It is obvious
that $Re(k)$ is the propagation factor and $Im(k)$ is the
attenuation factor for a time harmonic plane wave with fixed angular
frequency $\omega$ in a dispersive material. Using the values of
$k$, the phase velocity ($v_p=\frac{\omega}{k_1}$) and the group
velocity ($v_g=\frac{d\omega}{dk_1}$) \cite{O} of the waves can be
calculated. The three values of the dispersion relation are shown in
Figs. 1, 2 and 3.

\begin{figure}
\center \epsfig{file=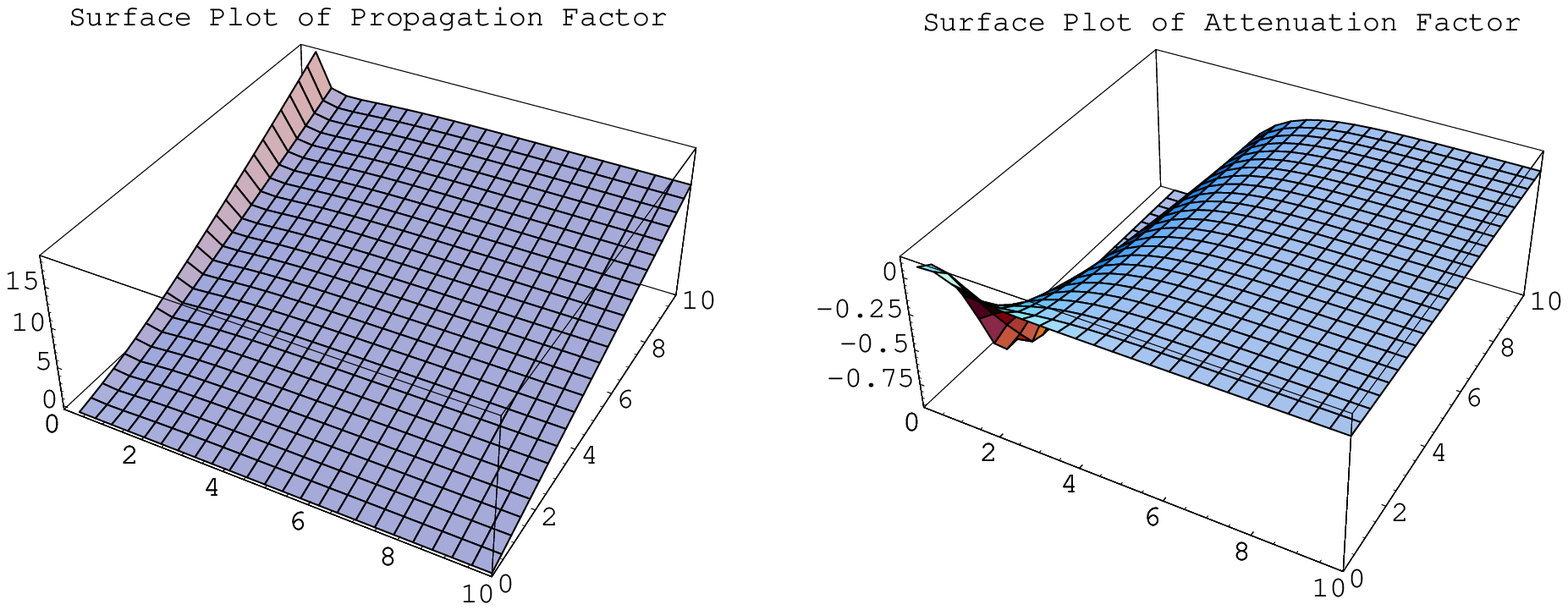,width=0.725\linewidth} \center
\epsfig{file=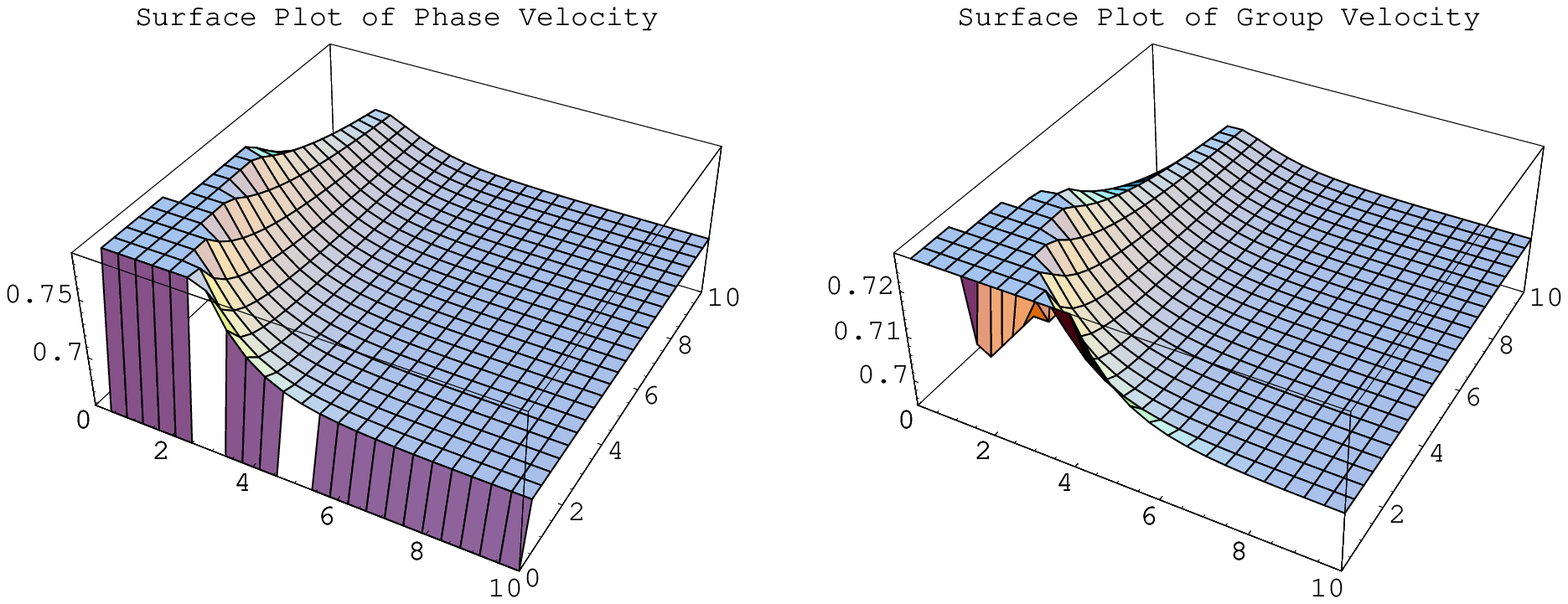,width=0.725\linewidth} \caption{Waves propagate
more near the event horizon. They damp as they move away from the
event horizon. Anomalous dispersion is found in that region.}
\end{figure}
Fig. 1 shows that the propagation factor is infinite at the horizon,
which indicates that no wave exists there. This has greater value
near the event horizon and decreases as we go away from the horizon.
It also increases with increasing angular frequency of the wave. The
attenuation factor increases, which shows that the waves damp with
increasing angular frequency and value of $z$. The phase velocity is
greater than the group velocity, which shows that dispersion is
normal in the region. The waves with negligible angular frequency
have phase velocities less than the group velocities and hence, the
medium shows anomalous dispersion.

\begin{figure}
\center \epsfig{file=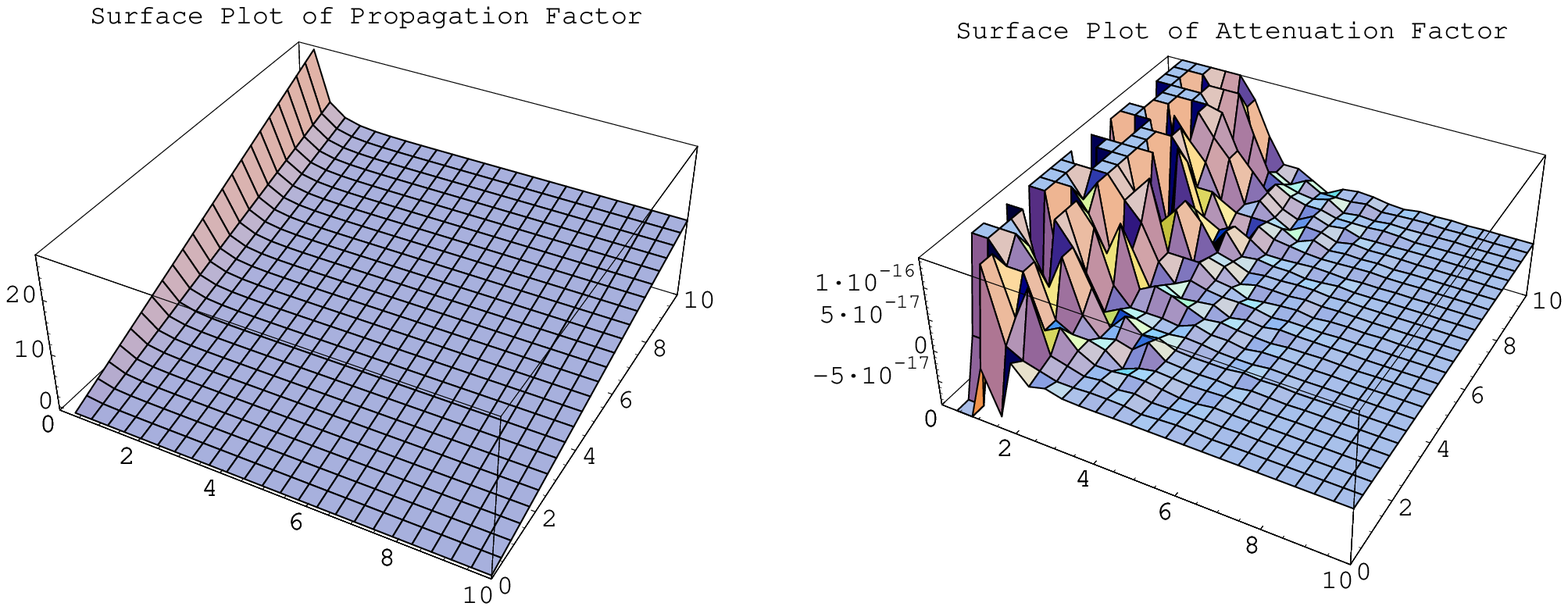,width=0.725\linewidth} \center
\epsfig{file=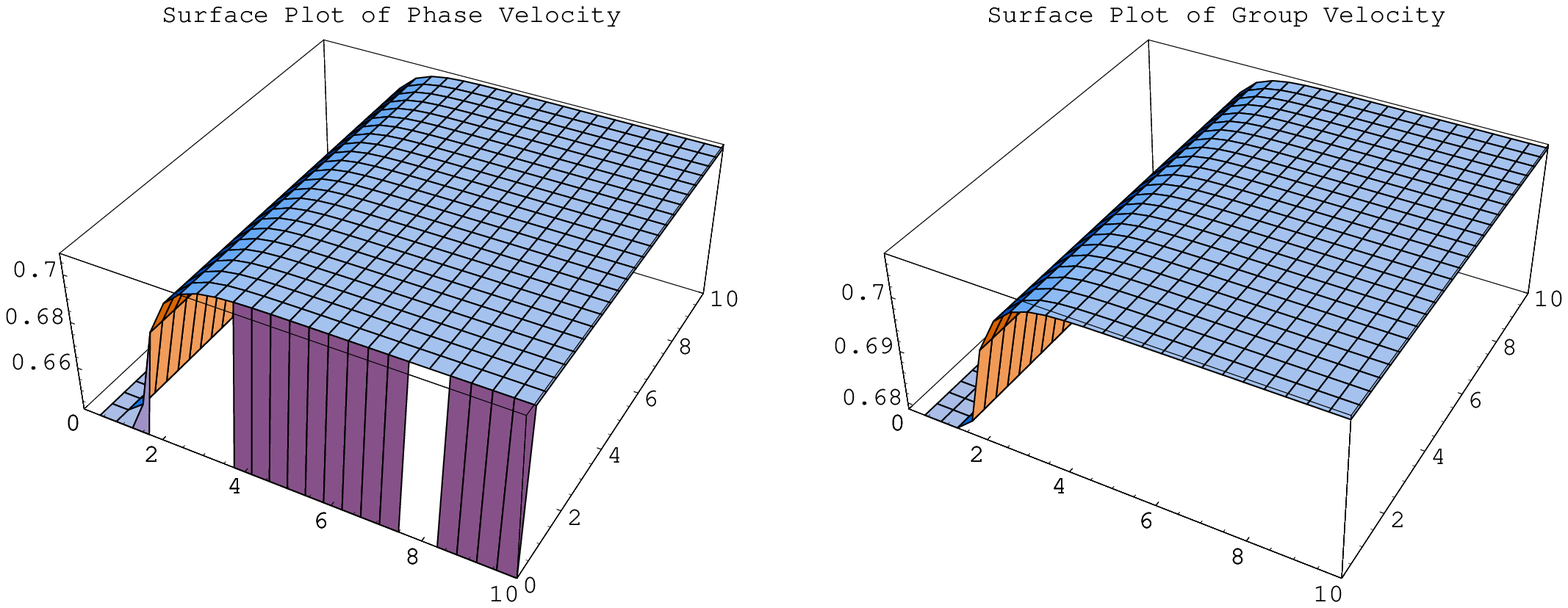,width=0.725\linewidth} \caption{Waves propagate
rapidly as they move near the event horizon. The dispersion of
waves is anomalous in that region.}
\end{figure}
The propagation factor in Fig. 2 shows the same behavior as that
in Fig. 1. In the region, the attenuation factor takes very small
values, which are positive and negative randomly. The phase and
the group velocities are the same in the region, except for zero
angular frequency waves. For those waves, the phase velocity is
less than the group velocity, indicating anomalous dispersion.

\begin{figure}
\center \epsfig{file=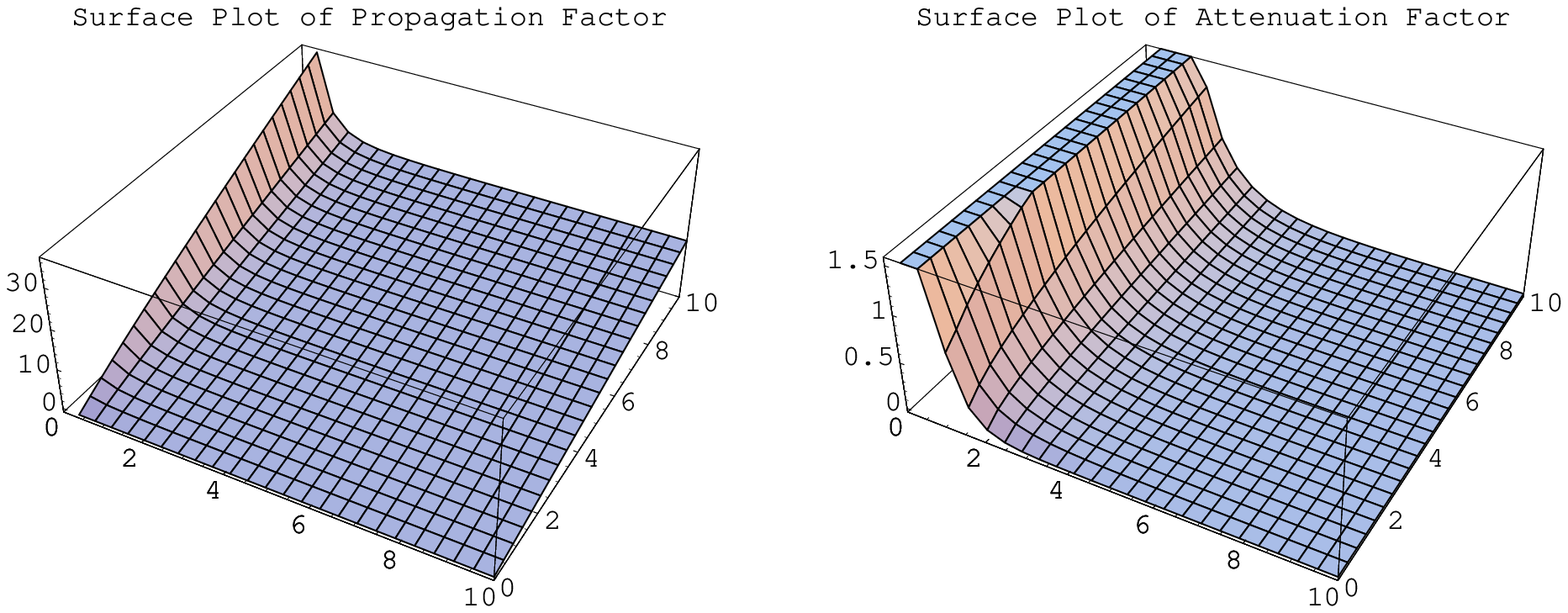,width=0.725\linewidth} \center
\epsfig{file=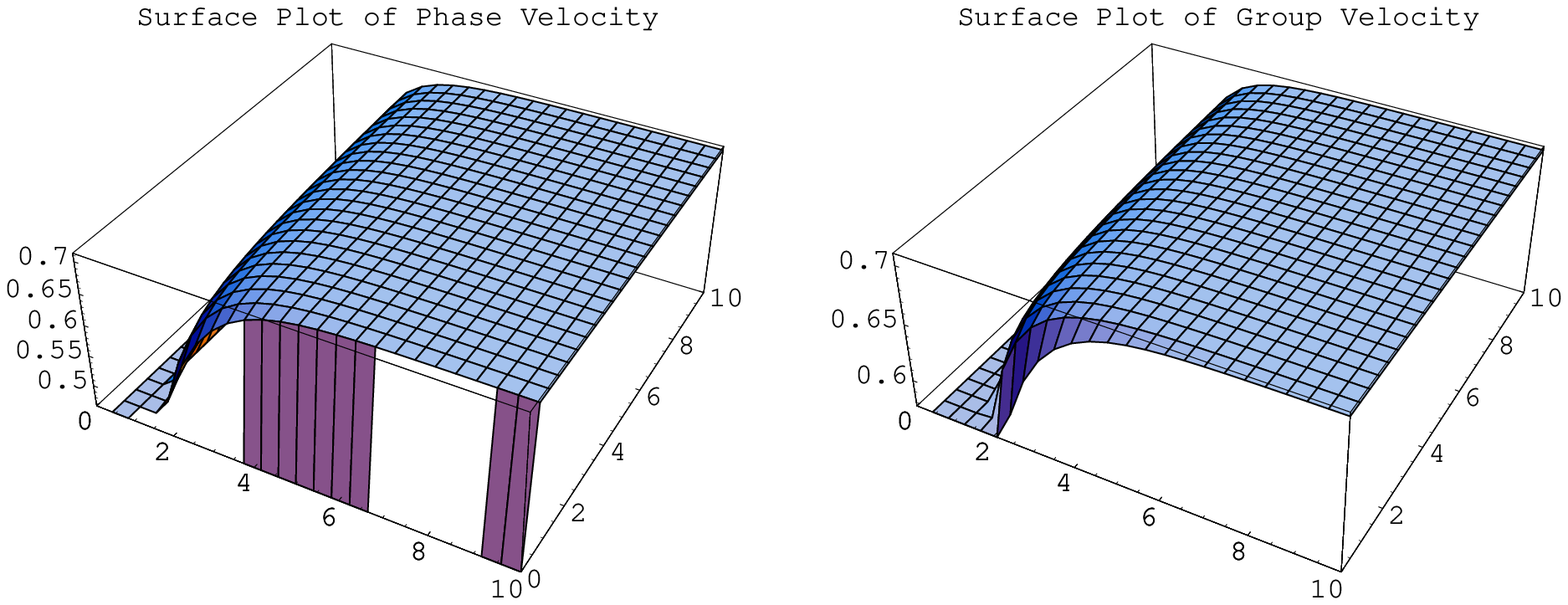,width=0.725\linewidth} \caption{Propagation of
waves is more near the event horizon. The waves grow as they move
away from the event horizon. The dispersion is found to be
anomalous.}
\end{figure}
In Fig. 3, the propagation factor is greater near the event horizon
and decreases as we go away from the horizon, but becomes infinite
at the event horizon. The attenuation factor increases when the
angular frequency increases and decreases when the value of $z$
increases. The waves grow as they go far from the event horizon of
the black hole. Damping occurs when the angular frequency increases.
The group velocity of the waves is greater than the phase velocity,
which shows anomalous dispersion.

\section{SUMMARY}

This paper is devoted to discussing the isothermal plasma wave
properties of the Schwarzschild black-hole's magnetosphere by
using the 3+1 formalism. For this purpose, the GRMHD equations are
derived and discussed by taking one-dimensional perturbations in
perfect MHD flow with its planar analogue. These equations are
written in component form and then Fourier analyzed by using the
assumption of plane waves. We have assumed that the black hole's
gravity does not introduce any effect on magnetic field. The
determinant of the coefficients are solved to get complex
dispersion relations, which yield the wave numbers. The properties
of plasma are inferred on the basis of this number, and the
relevant quantities are obtained in graphical form. A summary of
the results is given below:

From the three dispersion relations, we find that the wave
propagation factor becomes infinite. This implies that the wave
number is infinite at the event horizon; hence, waves vanish
there. This corresponds to the well-known fact that no information
can be extracted from a black hole.

The attenuation factor assumes extreme values near the event
horizon. These values gradually decrease or increase as they move
away from the horizon. The gravity of the black hole causes the
attenuation factor to take extreme values. Thus, the wave growth or
damping is larger near the event horizon. The attenuation factor is
indefinite at the event horizon. The wave propagation factor is high
near the black hole's event horizon. As the waves go far from the
event horizon, the propagation decreases, which shows that the
perturbations are high near the gravitational well.

It is worthwhile mentioning here that waves are highly excited near
the event horizon (Fig. 1). This excitement decreases as they move
away from the event horizon. This is due to immense gravity of the
black hole near the event horizon. Similar is the case of Fig. 2,
where the waves are randomly excited near the event horizon.

We notice that the dispersion is anomalous here while for a cold
plasma (Fig. 2) \cite{U}, normal wave dispersion was found. This
indicates that the plasma pressure causes the waves to pass
through the region in the neighborhood of the event horizon.

The dispersion relations are obtained by using the 3+1 ADM formalism
and contain the factor of acceleration (depends on the lapse
function and is equal to $-g$), which makes them different from the
usual MHD dispersion relations. We have investigated the waves
propagating in a plasma influenced by the gravitational field. We
observe that the values of $k$ show gravitational effects on the
smooth harmonic wave type perturbations. It would be interesting to
investigate the plasma wave properties by taking a rotating
background.

\vspace{0.5cm}

\begin{center}{\bf ACKNOWLEDGMENT}\end{center}

\vspace{0.5cm}

We appreciate the Higher Education Commission, Islamabad,
Pakistan, for its financial support during this work through the
{\it Indigenous PhD 5000 Fellowship Program Batch-II}.

\end{document}